\documentclass[twocolumn,superscriptaddress,amsmath,amssymb,pra,longbibliography]{revtex4-1}

\ifx\pdfoutput\@undefined\usepackage[usenames,dvips]{color}
\else\usepackage[usenames,dvipsnames]{color}
\usepackage{graphicx}
\usepackage{bm}
\usepackage{amssymb}
\usepackage{hyperref}
\usepackage{color}

\usepackage{amsmath}

\usepackage[T1]{fontenc}
\usepackage[latin1]{inputenc}

\newcommand{\CKB}{\textcolor{black}}
\newcommand{\CDS}{\textcolor{black}}

\usepackage{graphicx}
\usepackage{color}

\begin{document}

\title{Knotted Polarizations and Spin in 3D Polychromatic Waves}

\author{Danica Sugic}
\affiliation{Theoretical Quantum Physics Laboratory, RIKEN Cluster for Pioneering Research, Wako-shi, Saitama 351-0198, Japan}
\affiliation{School of Physics and Astronomy, University of Birmingham, Birmingham B15 2TT, United Kingdom}

\author{Mark R. Dennis}
\affiliation{School of Physics and Astronomy, University of Birmingham, Birmingham B15 2TT, United Kingdom}
\affiliation{EPSRC Doctoral Training Centre for Topological Design, University of Birmingham, Birmingham B15 2TT, United Kingdom}

\author{Franco Nori}
\affiliation{Theoretical Quantum Physics Laboratory, RIKEN Cluster for Pioneering Research, Wako-shi, Saitama 351-0198, Japan}
\affiliation{Physics Department, University of Michigan, Ann Arbor, Michigan 48109-1040, USA}

\author{Konstantin Y. Bliokh}
\affiliation{Theoretical Quantum Physics Laboratory, RIKEN Cluster for Pioneering Research, Wako-shi, Saitama 351-0198, Japan}


\begin{abstract}
We consider complex 3D polarizations in the interference of several vector wave fields with different commensurable frequencies and polarizations. We show that the resulting polarizations can form {\it knots}, and interfering three waves is sufficient to generate a variety of Lissajous, torus, and other knot types. We describe the spin angular momentum, generalized Stokes parameters and degree of polarization for such knotted polarizations, which can be regarded as partially-polarized. Our results are generic for any vector wave fields, including, e.g., optical and acoustic waves. As a directly-observable example, we consider knotted trajectories of water particles in the interference of surface water (gravity) waves with three different frequencies.
\end{abstract}

\maketitle

\section{Introduction}

Polarization is a fundamental property of vector waves of different nature. It is thoroughly studied in optics and electromagnetism \cite{Azzam_book}, but can be equally applied to any vector wavefields, e.g., elastic and acoustic waves \cite{Auld_book,Shi2019,Bliokh2019_II}. Polarization can be associated with the curve traced by the field vector $\bm{\mathcal F}({\bf r},t)$ in a given point ${\bf r}$.
For a monochromatic 3D wave field $\bm{\mathcal F}({\bf r},t) = {\rm Re} \left[ {\bf F}({\bf r}) e^{-i\omega t}\right]$,
this curve is generically an ellipse \cite{Azzam_book}. 

Rotation of the field vector at a given point ${\bf r}$ can also be associated  with an intrinsic angular momentum (AM), i.e., {\it spin} \cite{Andrews_book,BerryDennis2001,Bliokh2015PR,Long2018,Shi2019,Bliokh2019_II,Burns2020}. This is one of the fundamental dynamical properties of vector waves, including electromagnetic, elastic, and acoustic ones. 
In the most general case, a 2D polarization state is described by the four Stokes parameters (one of which is the normal spin component) \cite{Azzam_book,Berry1998}, while 3D polarized fields require nine generalized Stokes parameters \cite{Carozzi2000,Setala2002,Dennis2004,Sheppard2014} (three of which are responsible for the spin components \cite{Dennis2004,Eismann2020}).

When the wave contains multiple frequencies, the motion of the field vector becomes more complicated and in the limit of an irregular chaotic-like motion implies total {\it depolarization} of the wave. However, when a polychromatic field contains only several frequency components with well-defined polarizations, the field vector motion is complicated but still regular. This regime is only barely studied; the only properly-described case involves 2D paraxial fields with two commensurable frequencies, which generate closed Lissajous-like polarization curves \cite{Freund2003_I,Freund2003_II,Fleischer2014,Pisanty2019}. 
\CDS{Notably, optical nonlinear wave-mixing processes essentially involve generic polarizations and spin angular momentum of all the harmonics \cite{Fleischer2014}. Therefore, there is strong motivation to take full control of the nontrivial 3D polarizations of polychromatic fields with commensurable frequencies.}

In this work, we show that interfering three or more waves in 3D with commensurable frequencies produces closed polarization `trajectories' which are generally {\it knotted} \cite{Rolfsen_book} \CDS{(i.e., cannot be continuously transformed into a circle without cuts or self-crossings)}. 
Knotted structures in wave fields have been intensively studied recently in various contexts \cite{Kauffman_book}, such as \CKB{knotted electromagnetic field lines \cite{Ranada1992,Irvine2008,Kedia2013,Arrayas2017}, knots of wave singularities \cite{BerryDennis2001_II,Leach2004,Dennis2010,Laroque2018,Sugic2018,Larocque2020}, and 2D Lissajous-like polarizations with an extra synthetic dimension \cite{Pisanty2019}.} However, the 3D `knotted polarizations' considered here is a novel physical entity, to the best of our knowledge. We will describe several classes of knots, which are naturally generated in polychromatic fields. We will also analyze the spin angular momentum and generalized Stokes parameters produced by such knotted polarizations. 
\CDS{These properties are highly important in contexts of light-matter and spin-orbit interactions involving high-harmonic generation \cite{Fleischer2014,Tancogne2017,Tang2020}.}
Our approach is general and can be applied to optical, acoustic, and other vector wave fields. In particular, we show an example of knotted polarizations in interference of surface water (gravity) waves. There, such polarizations correspond to {\it real-space trajectories} of water molecules and can be directly observed experimentally \cite{Francois2017}.

\section{Knotted polarizations}\label{biorthogonal}

\begin{figure*}[t!]
{\centering \includegraphics[width=\textwidth]{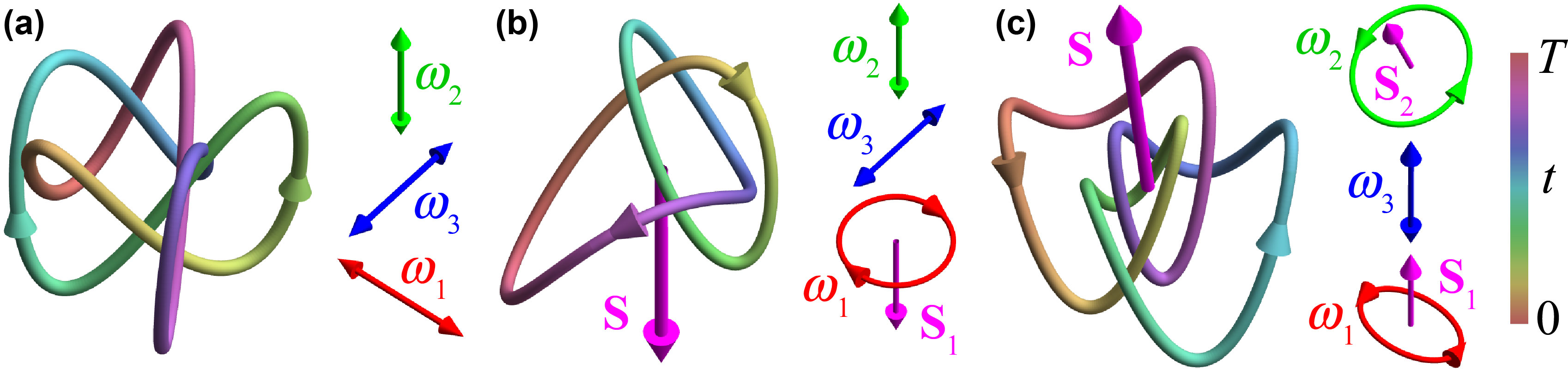}}
\caption{Examples of knotted polarizations $\bm{\mathcal F}(t)$ in the interference of three waves with different frequencies and polarizations (shown to the right from the knots). (a) The Lissajous knot (\ref{Eq.1}) with $A_1=A_2=A_3$, $\omega_2/\omega_1 = 5/2$, $\omega_3/\omega_1 = 3/2$, $\phi_1 - \phi_2 = 0.04$, and $\phi_3 - \phi_2 = 1.8$ (the $5_2$ or three-twist knot \cite{Dennis2017}). (b) The torus knot (\ref{Eq.2}) with $p=2$, $q=3$ (trefoil knot). (c) Figure-eight knot (\ref{Eq.3}). The spins of the interfering waves and resulted knotted polarizations are shown in purple, whereas the 3D degree of polarization for the knotted states (a)--(c) are $P=0$, $P=1/2$, and $P\simeq 0.67$, respectively \cite{SM}. 
\label{Fig1}}
\end{figure*}

We consider an interference of 3D vector fields with multiple commensurable frequencies $\omega_n$, $n=1,...,N$, so that 
$\bm{\mathcal F}({\bf r},t) = \sum_{n} {\rm Re}  \left[ {\bf F}_n({\bf r}) e^{-i\omega_n t}\right] 
$. The field vector in a given point ${\bf r}$ traces a closed 3D curve with the temporal period $T=2\pi/\Omega$, where $\Omega$ is the lowest common multiple of $\{\omega_n\}$.
Such polarization curves can be topologically nontrivial and form {\it knots} \cite{Rolfsen_book}, see Fig.~\ref{Fig1}.
Note that the polarization curve depends only on the frequencies and elliptical polarizations of the interfering waves ${\bf F}_n$ in the given point ${\bf r}$. From now on we fix this point and only consider temporal dependencies of the fields. 

Because of the time-harmonic character of the interfering waves, 
the Cartesian field components $\bm{\mathcal F} (t) = \left[ {\mathcal F}_x (t), {\mathcal F}_y (t), {\mathcal F}_z (t) \right]$ are sums of $\cos(\omega_n t)$ and $\sin(\omega_n t)$ terms with different amplitudes. Knotted curves described by such harmonic terms are known as {\it Fourier or harmonic knots} \cite{Trautwein1995,Kauffman1997}. These knots are labeled by three integer indices $(i,j,k) \geq (1,1,1)$ indicating the numbers of frequencies in the three Cartesian field components. 
\CKB{Remarkably, any type of knot can be constructed as a Fourier-$(1,1,k)$ knot with some $k$ \cite{Lamm1999}.}

In the simplest case, when each Cartesian component has only one frequency, such Fourier-$(1,1,1)$ knots are called {\it Lissajous knots} \cite{Bogle1994,Lamm1997,Boocher2009}, with $\bm{\mathcal F} (t)$ given by: 
\begin{align}
\label{Eq.1}
\left[
A_1 \cos(\omega_1 t + \phi_1), A_2 \cos(\omega_2 t + \phi_2), A_3 \cos(\omega_3 t + \phi_3)
\right], 
\end{align}
where the frequencies $\omega_n$, $n=1,2,3$, are proportional to three coprime integers, whereas $A_n$ and $\phi_n$ are amplitudes and phases. In physical terms, the Lissajous-knot polarizations are produced by a superposition of {\it three  linear polarizations} oriented along the three Cartesian axes, as shown in Fig.~\ref{Fig1}(a) for the example of $5_2$ or `three-twist' knot \cite{Rolfsen_book,Kauffman_book,Lamm1997,Dennis2017}. 

One of the most important classes of knots are {\it torus knots} ${\rm T}_{p,q}$, which lie on a torus surface and are characterized by a pair of coprime integers $(p,q)$ \cite{Rolfsen_book,Kauffman_book}. Notably, every type of torus knots can be represented by a Fourier-$(1,1,2)$ knot with $\bm{\mathcal F} (t)$ given by \cite{Hoste2007} 
\begin{align}
\label{Eq.2}
&A \left[ \cos(\omega_1 t),   
\cos\!\left(\omega_2 t+\phi_2\right) ,  
-\sin(\omega_1 t) + \cos\!\left( \omega_3 t + \phi_3 \right)
\right], \nonumber \\
&\frac{\omega_2}{\omega_1} = \frac{q}{p},~~\frac{\omega_3}{\omega_1} = \frac{q-p}{p},~~ 
 \phi_2 = \frac{\pi}{2p},~~\phi_3 = \frac{\pi}{2p} - \frac{\pi}{4q}\,. 
\end{align}
Polarization torus knots are produced by a superposition of {\it one circular and two linear polarizations}, as shown in Fig.~\ref{Fig1}(b) for the example of trefoil knot ${\rm T}_{2,3}$ \cite{Rolfsen_book,Kauffman_book}. Note that waves with frequencies satisfying Eq.~(\ref{Eq.2}) can be generated in nonlinear wave-mixing processes \cite{Boyd_book,Pisanty2019}.

The Lissajous and torus knots do not exhaust all possible Fourier knots. For example, the {\it figure-eight knot} can be generated by the field $\bm{\mathcal F} (t)$ as follows \cite{Kauffman1997}:
\begin{align}
\label{Eq.3}
& A \left[ \cos(\omega t) + \cos(3 \omega t),   
0.4 \sin (3 \omega t) - \sin(6 \omega t), \right. \nonumber \\  
& \left. 0.6 \sin(\omega t) + \sin(3 \omega t) \right]. 
\end{align}
This is a superposition of {\it two elliptical and one linear polarizations}, Fig.~\ref{Fig1}(c). 

\section{Spin in optical and acoustic fields}\label{biorthogonal}

The normalized period-averaged spin AM density in a monochromatic vector wavefield can be characterized by the expression ${\bf S} = {\rm Im}\, ({\bf F}^*\! \times {\bf F})/|{\bf F}|^2$, $|{\bf S}| \leq 1$. In optical fields, ${\bf F} = {\bf E}$ is the electric field \cite{Berry1998,BerryDennis2001,BAD2019}, while in sound wavefields in fluids or gases, ${\bf F} = {\bf V}$ is the velocity field \cite{Shi2019,Bliokh2019_II,Burns2020}. Note that this simplified approach ignores the presence of other fields (magnetic field in optics and pressure field in acoustics \cite{Andrews_book,Bliokh2015PR,Berry2009,Cameron2012,Bliokh2013,Bliokh2019_II,Burns2020}), but this omission is justified in many practical problems, where experimental measurements and phenomena are sensitive only to the electric and velocity fields. 

The above expression for the normalized spin originates from the time-averaged non-normalized expression $\langle \bm{\mathcal G} \times \bm{\mathcal F} \rangle = {\rm Im}\, ({\bf F}^*\! \times {\bf F})/(2\omega)$ and the density of ``field quanta'' given by the energy density divided by frequency, $\langle \bm{\mathcal F} \cdot \bm{\mathcal F} \rangle/\omega = |{\bf F}|^2/(2\omega)$, where $\bm{\mathcal G}$ is the vector-potential, such that $\bm{\mathcal F} = \partial_t \bm{\mathcal G}$. In optics, $\bm{\mathcal A} = - \bm{\mathcal G}$ is the magnetic vector-potential in the Coulomb gauge \cite{Bliokh2013}, while in acoustics $\bm{\mathcal R} = \bm{\mathcal G}$ is the displacement field \cite{Burns2020}, so that $\bm{\mathcal G} \times \bm{\mathcal F} = \bm{\mathcal R} \times \bm{\mathcal V}$ is the natural mechanical AM form. 

In the polychromatic field $\bm{\mathcal F}$ considered above, the vector-potential equals $\bm{\mathcal G}({\bf r},t) = \sum_{n} \omega_n^{-1} {\rm Im}  \left[ {\bf F}_n({\bf r}) e^{-i\omega_n t}\right]$. Substituting it into the spin form $\langle \bm{\mathcal G} \times \bm{\mathcal F} \rangle$ with time-averaging over the period $T$ yields the normalized spin density in a polychromatic field:
\begin{equation}
\label{Eq.4}
{\bf{S}} = \frac{{\sum_n {\omega _n^{ - 1}{{\rm Im}} \left( {{\bf{F}}_n^* \times {{\bf{F}}_n}} \right)} }}{{\sum_n {\omega _n^{ - 1}{{\left| {{{\bf{F}}_n}} \right|}^2}} }}\,. 
\end{equation}
This equation shows that the spin of a polychromatic field represents a properly weighted sum of the spins ${\bf S}_n$ of the interfering monochromatic components. In the particular case of 2D bi-chromatic optical fields, consisting of circularly-polarized waves, the general expression (\ref{Eq.4}) coincides with the one used in Refs.~\cite{Freund2003_I,Fleischer2014}. Obviously, the spin (\ref{Eq.4}) vanishes for interfering linearly-polarized fields, such as the Lissajous knots, Fig.~\ref{Fig1}(a). For other knotted polarizations it is generically nonzero (see \cite{SM} and Figs.~\ref{Fig1}(b,c) for the spin of the torus and figure-eight knotted polarizations, Eqs.~(\ref{Eq.2}) and (\ref{Eq.3})), and restricted by $|{\bf S}|\leq 1$.

Remarkably, the above spin AM in polychromatic fields with complex 3D polarization curves allows a very simple mechanical analogy. Let us consider a mechanical point particle of unit mass moving in real space along the closed trajectory ${\bf r}(t) = \bm{\mathcal G}(t)$. Then, the period-averaged mechanical AM of this particle, $\langle {\bf r} \times \partial_t {\bf r} \rangle$ equals the spin $\langle \bm{\mathcal G} \times \bm{\mathcal F} \rangle$, whereas its averaged kinetic energy, $\langle |\partial_t {\bf r}|^2/2 \rangle$ is half of the field energy $\langle \bm{\mathcal F} \cdot \bm{\mathcal F} \rangle$.
This hints that the curve traced by the vector-potential $\bm{\mathcal G}$ could be more fundamental than the one traced by the field $\bm{\mathcal F}$. Of course, all previous considerations about knotted polarizations can be equally applied to the vector-potential polarization.

Note that for monochromatic fields, the polarization ellipses of the field  $\bm{\mathcal F}$ and its vector-potential $\bm{\mathcal G}$ coincide with each other (up to a constant factor). In contrast, the complex polarization curves of polychromatic fields and their vector-potentials generally differ considerably. This is because single-harmonic elliptical motion is invariant with respect to the time-derivative operation, while complex polychromatic motion is not.

\section{Generalized Stokes parameters and depolarization}\label{biorthogonal}

The above consideration of the spin AM suggests that any {\it quadratic forms} of fields could be considered in a similar manner, such that interference terms between different frequencies are averaged out and the form represents a weighted sum of contributions from each frequency component. For example, the canonical momentum of a polychromatic field is calculated similarly to the spin but with the substitution of the quadratic form ${\rm Im} \left( {\bf F}_n^*\times {\bf F}_n \right) \to {\rm Im} \left[ {\bf F}_n^*\cdot (\bm{\nabla}) {\bf F}_n \right]$ \cite{Berry2009,Bliokh2015PR,Shi2019,Bliokh2019_II,Burns2020,Bliokh2013}. Calculation of optical or acoustic radiation forces and torques also involves similar quadratic forms and allows a similar approach \cite{Bliokh2014,Toftul2019}. 

Here we consider important quadratic forms used for the description of 3D partially polarized fields, namely, the 
\CKB{{\it generalized Stokes parameters} $\Lambda_l$, $l=0,1,...,8$ \cite{Carozzi2000,Setala2002,Dennis2004,Sheppard2014}. For monochromatic fields, these parameters appear from the Hermitian $3 \times 3$ coherence matrix $\Gamma_{ij} = \langle F_i^* F_j \rangle$, $i,j = x,y,z$, and its decomposition  $\Gamma_{ij} = \frac{1}{3}\sum_{l=0}^{8} \Lambda_l \{\lambda_l\}_{ij}$
via the Gell-Mann matrices ${\lambda}_l$ \cite{Carozzi2000,Setala2002,Dennis2004,Sheppard2014}.
Here} $\Lambda_0 = {\rm Tr} (\hat\Gamma)$ is associated with the field intensity, parameters $\Lambda_1$, $\Lambda_3$, $\Lambda_4$, $\Lambda_6$, and $\Lambda_8$ are related to the real part of $\hat\Gamma$, while the three parameters $\Lambda_7$, $\Lambda_5$, and $\Lambda_2$ related to the imaginary part of $\hat\Gamma$ are proportional to the Cartesian components of the spin density ${\bf S}$ \cite{Dennis2004,Eismann2020}. 

\CKB{For the polychromatic fields considered here, we introduce the natural analogues of the coherence matrix and the generalized Stokes parameters as:
\begin{align}
\Gamma_{ij} = \sum_n F_{ni}^* F_{nj}
= \frac{1}{3}\sum_{l=0}^{8} \Lambda_l \{\lambda_l\}_{ij},
\label{Eq.5}
\end{align}
such that the polarization parameters are sums of the corresponding parameters for the interfering monochromatic waves: $\Lambda_l = \sum_n \Lambda_{l}^{(n)}$ \cite{SM}. Note the the vector $2(-\Lambda_7,\Lambda_5,-\Lambda_2)/(3\Lambda_0 )$ represents an analogue of the normalized spin (\ref{Eq.4}) but without $\omega_n^{-1}$ weighting factors \cite{Dennis2004,Eismann2020}.}

Among the different definitions of the degree of polarization for 3D fields \cite{Ellis2005,Setala2002,Dennis2007,Pertruccelli2010}, a natural choice is $P = {\sqrt{\sum_{l=1}^{8} \Lambda_l^2}}\left/{\sqrt{3}\Lambda_0} \right. \in [0,1]$.
An elliptically polarized monochromatic light, i.e., each of the interfering components in our knotted fields, Eqs.~(\ref{Eq.1})--(\ref{Eq.3}) and Fig.~\ref{Fig1}, is fully polarized: $P^{(n)}=1$ \cite{SM}. However, calculating the polarization parameters (\ref{Eq.5}) for polychromatic knotted fields, we find that its degree of polarization diminishes: $P<1$. 
This means that polychromatic knotted fields can be regarded as {\it partially depolarized}. In particular, for any Lissajous-knotted field (\ref{Eq.1}) with $A_1=A_2=A_3$, we find that $P=0$, i.e., it is {\it totally unpolarized} \cite{SM}. This is natural, because such fields consists of three independent oscillations with equal amplitudes along the three axes. In turn, for any torus-knotted field (\ref{Eq.2}), the degree of polarization is $P=1/2$, while for the figure-eight-knotted polarization (\ref{Eq.3}) we find $P\simeq 0.67$ 
\cite{SM}. In these cases, the presence of partial polarization is related to the presence of nonzero spin (\ref{Eq.4}) in such knotted fields, Figs.~\ref{Fig1}(b,c) \cite{Eismann2020}; total depolarization implies zero spin.      

\CKB{Note, however, that the spin, polarization parameters, and degree of polarization are not directly related to {\it topological} properties of knotted polarizations. These are rather {\it geometrical} (or dynamical) properties of the field curve $\bm{\mathcal F}(t)$. For example, the polarization parameters of the Lissajous knots (\ref{Eq.1}) strongly depend on the amplitudes $A_n$ (scaling factors along the Cartesian axes) \cite{SM}, while the knot topology is obviously independent of these.}   

\begin{figure*}[t!]
{\centering \includegraphics[width=0.96\textwidth]{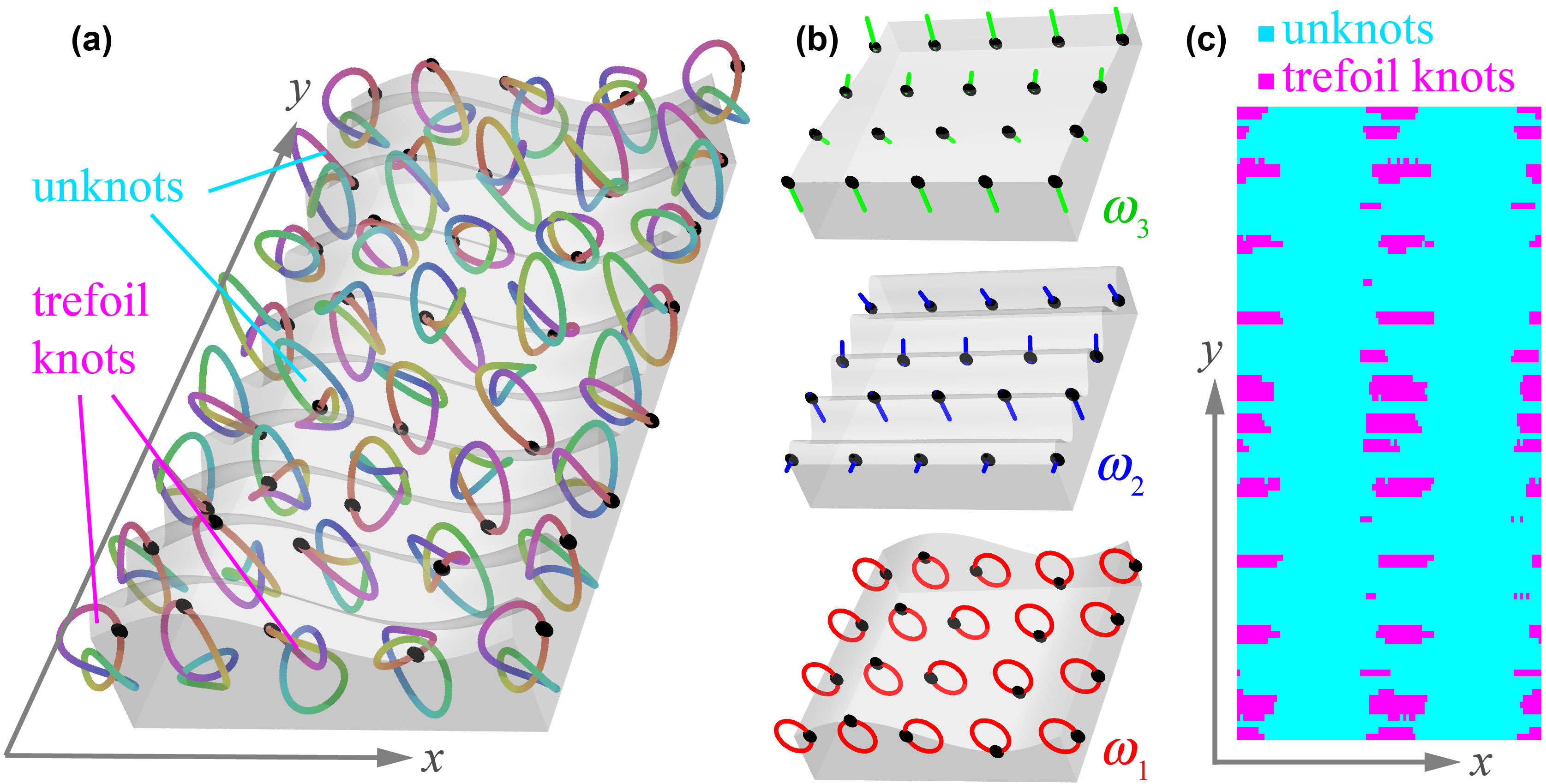}}
\caption{Interference of three surface water waves, one propagating in $x$ and two standing along $y$, Eq.~(\ref{Eq.6}), with different frequencies and phase corresponding to Fig.~\ref{Fig1}(b). The instantaneous surface shapes (greyscale) and 3D water-particles trajectories (colored) are shown for the interference field (a) (see also its animated version \cite{SM}) and each of the interfering waves (b). Some of these trajectories correspond to trefoil torus knots, Fig.~\ref{Fig1}(b), while other are `unknots'. The $(x,y)$-regions with knotted and unknotted trajectories are shown in (c).
\label{Fig2}}
\end{figure*}

\section{Knotted trajectories in water waves}\label{biorthogonal}

Remarkably, complex polarizations of polychromatic waves is a directly observable phenomenon. While measuring the time-dependent electric field with optical frequencies is practically impossible, the sound-wave polarization is related to the velocity field $\bm{\mathcal V}({\bf r},t)$ or the displacement `vector-potential' $\bm{\mathcal R}({\bf r},t)$ \cite{Burns2020}. Here, the time-dependent field $\bm{\mathcal R}({\bf r},t)$ describes the real-space displacement of the medium particles, so that its polarization curve is the {\it real-space trajectory} of the particle. Although it is a challenge to observe the displacement of air or water particles at typical sound frequencies, this can be easily done for {\it surface water waves} with much lower frequencies and directly-observable motion of water particles. Indeed, a recent experiment \cite{Francois2017} observed a 2D Lissajous-like motion in the interference of water waves with different frequencies. 
Here we show that a similar experiment can be designed to observe 3D knotted trajectories of water particles when taking into account their horizontal and vertical motions and interfering three waves with different frequencies.

Using equations of hydrodynamics, the equation of motion for the 3D displacement field $\bm{\mathcal R} ({\bf r}_\bot,t)= \left[ {\mathcal X}({\bf r}_\bot,t), {\mathcal Y}({\bf r}_\bot,t), {\mathcal Z}({\bf r}_\bot,t)\right]$ on the water surface $z=0$ for deep-water (gravity) waves can be written as \cite{SM} $\partial_t^2 {\mathcal Z} = g\, \bm{\nabla}_\bot \cdot \bm{\mathcal R}_\bot$, where ${\bf r}_\bot = (x,y)$, $\bm{\nabla}_\bot = (\partial_x,\partial_y)$, $\bm{\mathcal R}_\bot = ({\mathcal X},{\mathcal Y})$, and
$g$ is the gravitational acceleration. 
For a monochromatic plane-wave field $\bm{\mathcal R} ({\bf r}_\bot,t) = {\rm Re} \left[ {\bf R}\, e^{-i\omega t +i {\bf k}\cdot {\bf r}_\bot} \right]$, with ${\bf k} = (k_x,k_y)$ being the wavevector, this yields $-\omega^2 Z = i g\, {\bf k}\cdot {\bf R}_\bot$, where $\omega^2 = g k$ is the dispersion relation \cite{LLfluid}. These equations describe a well-known fact: in a plane gravity wave, the water particles move along circular trajectories lying in the plane determined by the wavevector and normal to the surface \cite{surfacewave}. In other words, a single deep-water wave has a purely {\it circular polarization}. Then, interfering oppositely-propagating waves with the same frequency yields a superposition of opposite circular polarization, i.e., a {\it linear polarization} with the position-dependent orientation. Thus, propagating and standing gravity waves provide circular and linear polarizations necessary for the torus knots (\ref{Eq.2}), Fig.~\ref{Fig1}(b). 

Explicitly, we consider the interference of an $x$-propagating plane wave and two standing waves along the $y$-axis with three different frequencies and phases but equal amplitudes. This results in the displacement field 
%
\begin{widetext}
\begin{equation}
\label{Eq.6}
\bm{\mathcal R} ({\bf r}_\bot,t) \propto \left( {\begin{array}{*{20}{c}}
{\cos \left( {{\omega _1}t - {k_1}x} \right)}\\
0\\
{ - \sin \left( {{\omega _1}t - {k_1}x} \right)}
\end{array}} \right) + 
\left( {\begin{array}{*{20}{c}}
0\\
{\cos \left( {{\omega _2}t + {\phi _2}} \right)\cos \left( {{k_2}y} \right)}\\
{\cos \left( {{\omega _2}t + {\phi _2}} \right)\sin \left( {{k_2}y} \right)}
\end{array}} \right) + \left( {\begin{array}{*{20}{c}}
0\\
{ - \cos \left( {{\omega _3}t + {\phi _3}} \right)\sin \left( {{k_3}y} \right)}\\
{\cos \left( {{\omega _3}t + {\phi _3}} \right)\cos \left( {{k_3}y} \right)}
\end{array}} \right),
\end{equation}
\end{widetext}
where $k_{1,2,3} = \omega_{1,2,3}^2 /g$. Choosing the frequencies and phases satisfying Eqs.~(\ref{Eq.2}), we find that the real-space trajectory of the water motion at ${\bf r} = {\bf 0}$, $\bm{\mathcal R} ({\bf 0},t)$, is exactly the torus knot (\ref{Eq.2}), Fig.~\ref{Fig1}(b).

Figure~\ref{Fig2} shows the water surface shape and water-particles trajectories $\bm{\mathcal R} ({\bf r}_\bot,t)$ for the whole interference field (\ref{Eq.6}) with $p=2$, $q=3$, corresponding to the trefoil knot in Fig.~\ref{Fig1}(b), and also for each of the interfering waves. Note that different points of the water surface $(x,y)$ correspond to different mutual phases of the interfering waves. Therefore, water motions in different points have different 3D trajectories. In our case, these trajectories represent trefoil knots and `unknots' \cite{Rolfsen_book,Kauffman_book}. The $(x,y)$-regions with knotted and unknotted trajectories are shown in Fig.~\ref{Fig2}(c). Numerical calculations \cite{Taylor2017} show that the fraction of knotted trajectories here is about $12\%$. 
\CKB{Many of the polarization trajectories are nearly self-intersecting, which can make the precise knot topology hard to resolve. Such self-intersections occur at every transition here between unknot and trefoil knot, but the arcs {\it pass through} each other (i.e. not {\it reconnecting} like vortices \cite{Berry2012,Kleckner2016}). More complicated superpositions give rise to larger areas of knotting, with transitions between multiple knot types.}

\section{Conclusions}

We have studied polychromatic 3D vector waves with closed (periodic) field trajectories. In particular, we have found that interference of three or more vector waves with different frequencies can generate a variety of {\it knotted} 3D polarizations, including Lissajous, torus, and other knots. We have introduced a natural formalism for the spin angular momentum (including a simple mechanical analogy), generalized Stokes parameters, and other quadratic forms for polychromatic yet periodic 3D vector waves. This revealed the presence of nonzero spin and partial depolarization in generic knotted polarizations. Finally, using the generic character of our consideration, valid for any 3D vector wave fields, we have provided an example where knotted polarizations appear as directly-observable trajectories of water particles in the interference of surface-water (gravity) waves.  

These results provide a new natural application of the knot theory to wave physics, different from the previously-studied knots of field lines \cite{Ranada1992,Irvine2008,Kedia2013,Arrayas2017} or singularities \cite{BerryDennis2001_II,Leach2004,Dennis2010,Laroque2018,Sugic2018,Larocque2020}. 
Moreover, our work considerably extends earlier studies of complex polarizations in polychromatic fields, previously restricted to 2D bichromatic Lissajous-like polarizations \cite{Freund2003_I,Freund2003_II,Fleischer2014,Pisanty2019}, as well as theory and potential applications of the spin AM, so far mostly restricted to monochromatic waves \cite{Andrews_book,BerryDennis2001,Bliokh2015PR,Shi2019,Bliokh2019_II}. 
\CDS{Since complex 3D polarizations and polychromatic fields are highly important for light-matter interactions, nonlinear and ultrafast processes, quantum control, etc. \cite{Brixner2004,Fleischer2014,Tancogne2017,Pisanty2019,Tang2020},
one can expect that nontrivial topological and dynamical features of knotted polarization states will find numerous applications in complex wave systems. } 


\vspace*{0.2cm}
\begin{acknowledgments}
This work was partially supported by NTT Research, Army Research Office (ARO) (Grant No.~W911NF-18-1-0358), Japan Science and Technology Agency (JST) (via the CREST Grant No.~JPMJCR1676), Japan Society for the Promotion of Science (JSPS) (via the KAKENHI Grant No.~JP20H00134, and the grant JSPS-RFBR Grant No.~JPJSBP120194828), the Grant No. FQXi-IAF19-06 from the Foundational Questions Institute Fund (FQXi) (via the Grant No. FQXi-IAF19-06), and the EPSRC Centre for Doctoral Training in Topological Design (EP/S02297X/1).
\end{acknowledgments}

\bibliography{References_1}

\end{document}